\documentstyle[11pt,newpasp,epsf,psfig,twoside]{article} 
\def\edcomment#1{\iffalse\marginpar{\raggedright\sl#1\/}\else\relax\fi}
\reversemarginpar

\begin{document}
\title{Dynamics of the Central kpc in Barred Galaxies: Theory and Modeling}
\author{Isaac Shlosman\altaffilmark{1,2}}
\affil{Joint Institute for Laboratory Astrophysics, University of Colorado,
Campus Box 440, Boulder, CO 80309-0440, USA}

\altaffiltext{1}{JILA Visiting Fellow.}
\altaffiltext{2}{Permanent address: Department of Physics and Astronomy,
University of Kentucky, Lexington, KY 40506-0055.}

\begin{abstract}
The central kpc of barred galaxies exhibits a wealth of morphological
information on different
components with clear dynamical consequences. These include nuclear rings,
spirals, bars, and more. We argue that this morphology is driven by large-scale
stellar bars and analyze its consequences for gas dynamics and
the distribution of
star-forming regions. Specifically, we focus on gas flows in nested bar systems
and study their origin, as well as  the
gravitational decoupling of gaseous nuclear
bars with and without gas self-gravity. We find that  the gas response in nested
bars differs profoundly from that in single bars, 
 and that no offset dust lanes form
in the nuclear bars. 
\end{abstract}

\section{Introduction}

For years, disk galaxies have been studied with unresolved centers, called nuclei, 
of up to a few kpc in size. Recent progress in ground-based instruments and
the availability of the {\it Hubble Space Telescope} ({\it HST\/}) has
allowed for the first
time a meaningful analysis of central morphology and kinematics, both stellar
and gaseous. Although our knowledge of the inner regions of disk galaxies is 
clearly incomplete, certain patterns in their dynamical evolution and their
relationship to larger and smaller spatial scales have emerged. 
 
At least dynamically, the inner parts of disk galaxies can be defined
by the position of the inner Lindblad resonance(s) (hereafter ILRs), typically
at about 1~kpc from the center. This resonance between the bar pattern speed
and the precession rate of periodic orbits plays an important role in filtering
density waves, either stellar or gaseous, propagating between the bar corotation
radius and the center (e.g., Binney \& Tremaine 1987). It is usually
delineated by elevated star formation rates and the
concentration of molecular gas
in one or more nuclear rings. These rings exhibit a rich morphology and can
serve as cold gas reservoirs for fueling the central activity, stellar and
non-stellar.

It is a well-known property of rotationally supported gravitational systems that
any departure from axial symmetry facilitates angular-momentum transfer
due to gravitational torques across large radial distances. The action of
torques can be described in terms of non-local gravitational viscosity and
can be very efficient, when the departures from axisymmetry are substantial,
driving the dynamical and secular evolution of these systems. 
Numerous effects of
such evolution have been detected in barred galaxies, ranging
from making them
more centrally concentrated by driving the gas inwards, to reducing 
radial abundance gradients. In the absence of non-axisymmetric perturbations,
the dynamical evolution is expected to slow down dramatically.

The above effects are due to large-scale stellar bars and have been
extensively studied during the last two decades. Gravitational torques
and other effects of such bars are most prominent on scales of several kpc,
where the ratio of non-axisymmetric-to-radial force peaks. At smaller 
galactocentric distances
the influence of these bars decreases dramatically,
and within the central kpc, inside the ILR(s), the bar potential becomes more
axisymmetric. The question then is, What  is the prime factor driving evolution
in these regions,  where the effects of large-scale bars are weak?

Here we review the basic morphologies of the central kpc in {\sl barred}
galaxies and analyze the possible dynamical role of nuclear rings,
nuclear spirals
 and nuclear bars. This is done using high-resolution {\it HST\/}
and ground-based observations of the last few years. In most cases
these morphological features are secondary, in the sense that they have been
caused, directly or indirectly, by  large-scale bars.

\section{Nuclear Rings and Large-Scale Bars: The Global Connection}

The gas response to the gravitational torquing of the large-scale bar depends
mainly on the galactic mass distribution and the pattern speed of the bar,
which determine the positions of the major resonances in the disk.
For a wide range of bar properties, the gas response is characterized by
offset dust lanes delineating large-scale shocks, which are stationary in the
bar frame of reference and govern the gas flow in the bar (Prendergast
1962). Athanassoula (1992) has presented a quantitative analysis of the flow
pattern between the ILRs and the corotation and has shown that the existence of
one or two ILRs is necessary to produce shocks which are offset from the bar
major axis. In the absence of even one ILR, the shocks are centered on the
axis. The subsequent search in the RC3 catalogue (de Vaucouleurs et al. 1991)
has revealed only a couple of barred galaxies which exhibit centered shocks
(E. Athanassoula, private communication).  
The shape of the offset dust lanes depends on
the bar strength; stronger bars are accompanied by straight dust lanes without
star formation, while weaker bars show curved lanes with some star formation.
Gas enters the shocks (roughly) perpendicularly, and afterwards follows their
outline, leading to a strongly radial inflow towards the ILRs.

\begin{figure}
\plottwo{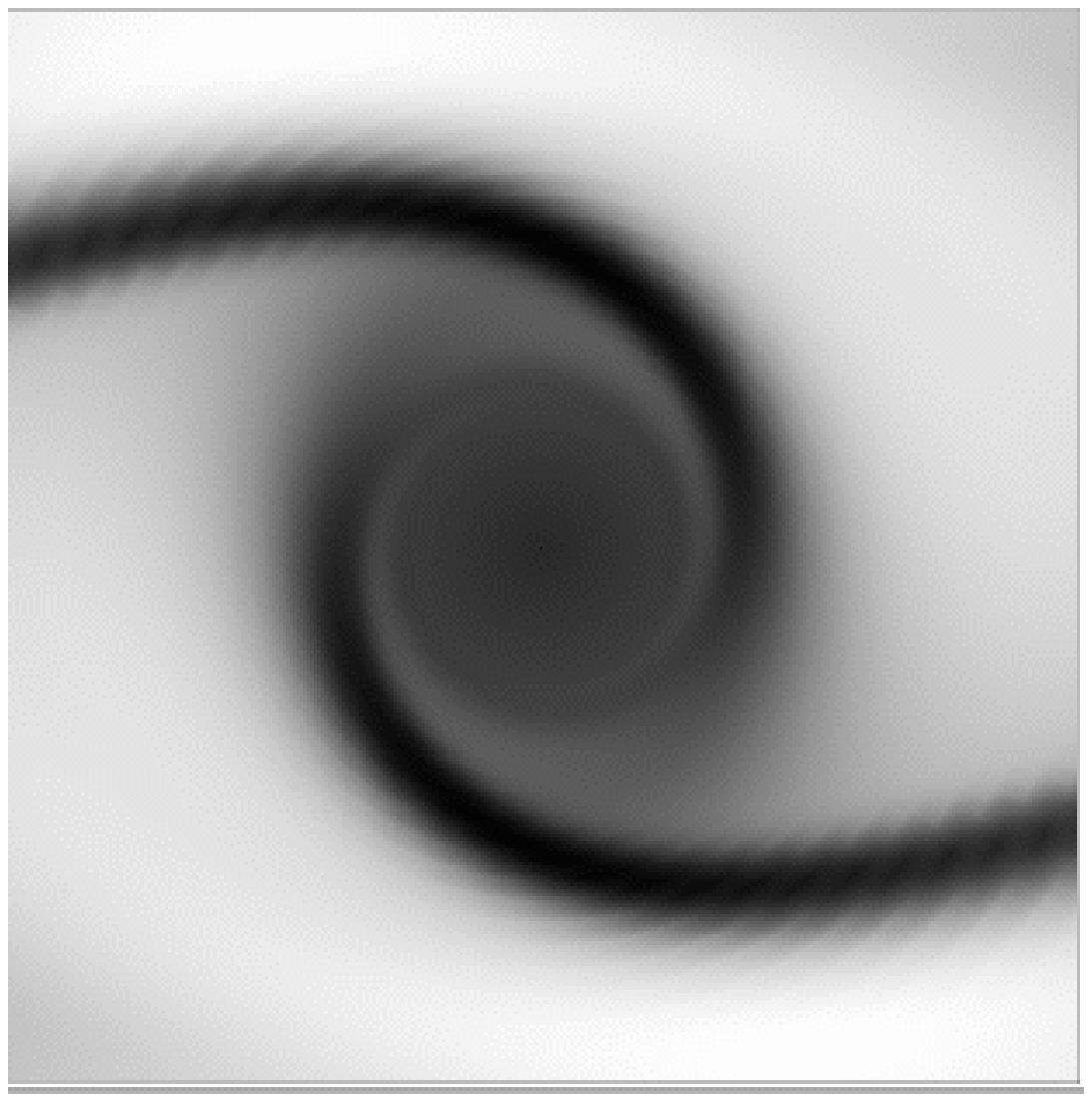}{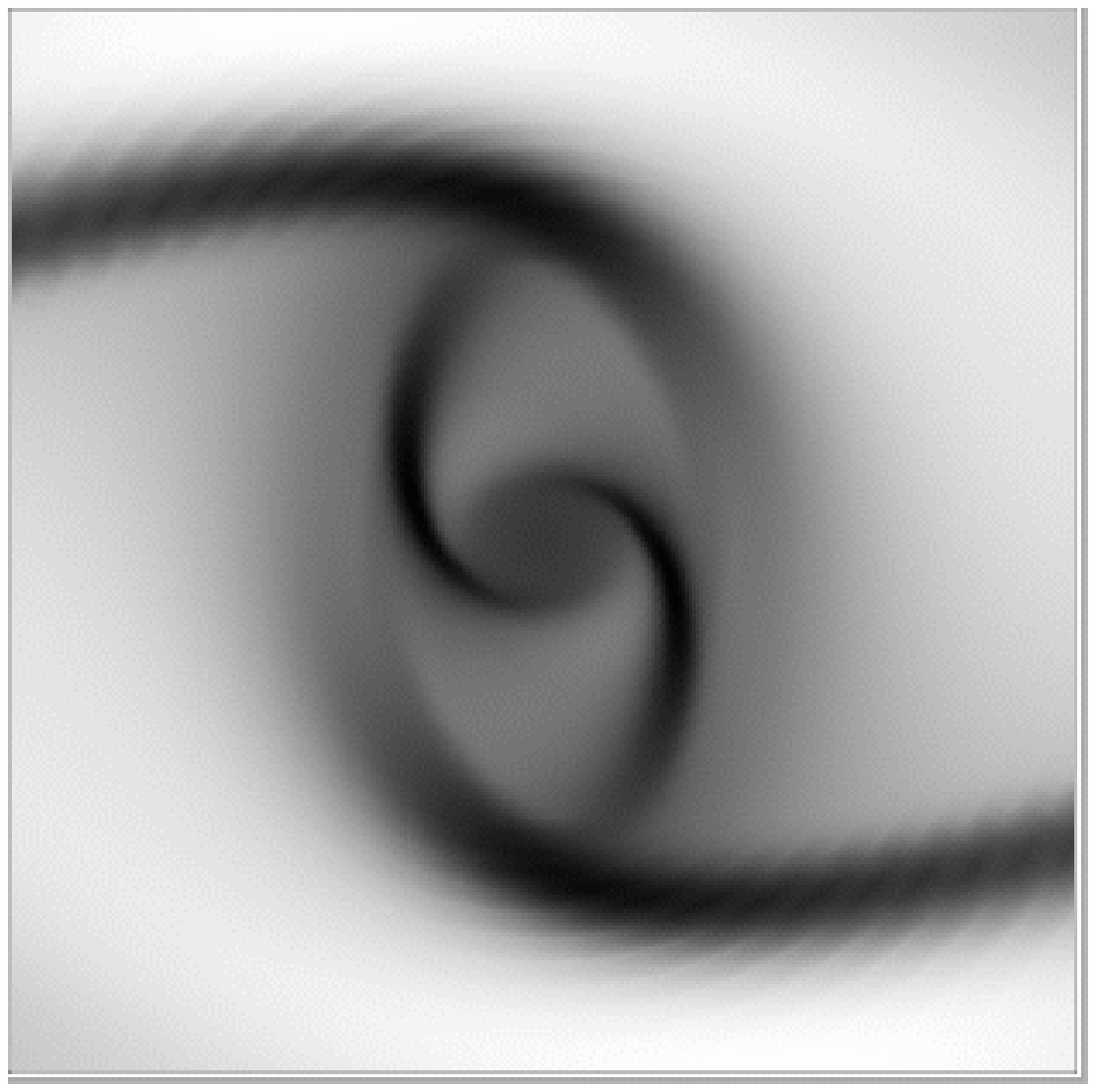}
\caption{{\it Grayscale} gas response to stellar bar torque. The bar is
horizontal and rotation is clockwise in its frame. Only the central resonance
region of 2~kpc is shown. {\it Left:\/} A single ILR is present causing a pair
of trailing spirals to form. The innermost gas settles on nearly circular $x_2$ 
orbits in an undisturbed nuclear disk. {\it Right:\/} A pair of ILRs is
present, causing two pairs of spirals to form, a leading  and a trailing
one. ({\it From Shlosman 1999.}) }
\end{figure}    

Across the ILR(s), shocks curve and become more circular, which results in a
shock-focused, azimuthal flow. Each ILR is characterized by a pair of shocks,
e.g., trailing at the outer ILR, leading at the inner ILR, and again trailing
at the nuclear Lindblad resonance (NLR), if one exists (Fig.~1). Reduction in
the radial inflow velocity leads naturally to gas accumulation at the above
radii, which form so-called nuclear rings.  

\begin{figure}[ht!!!!!!!!!!!]
\plottwo{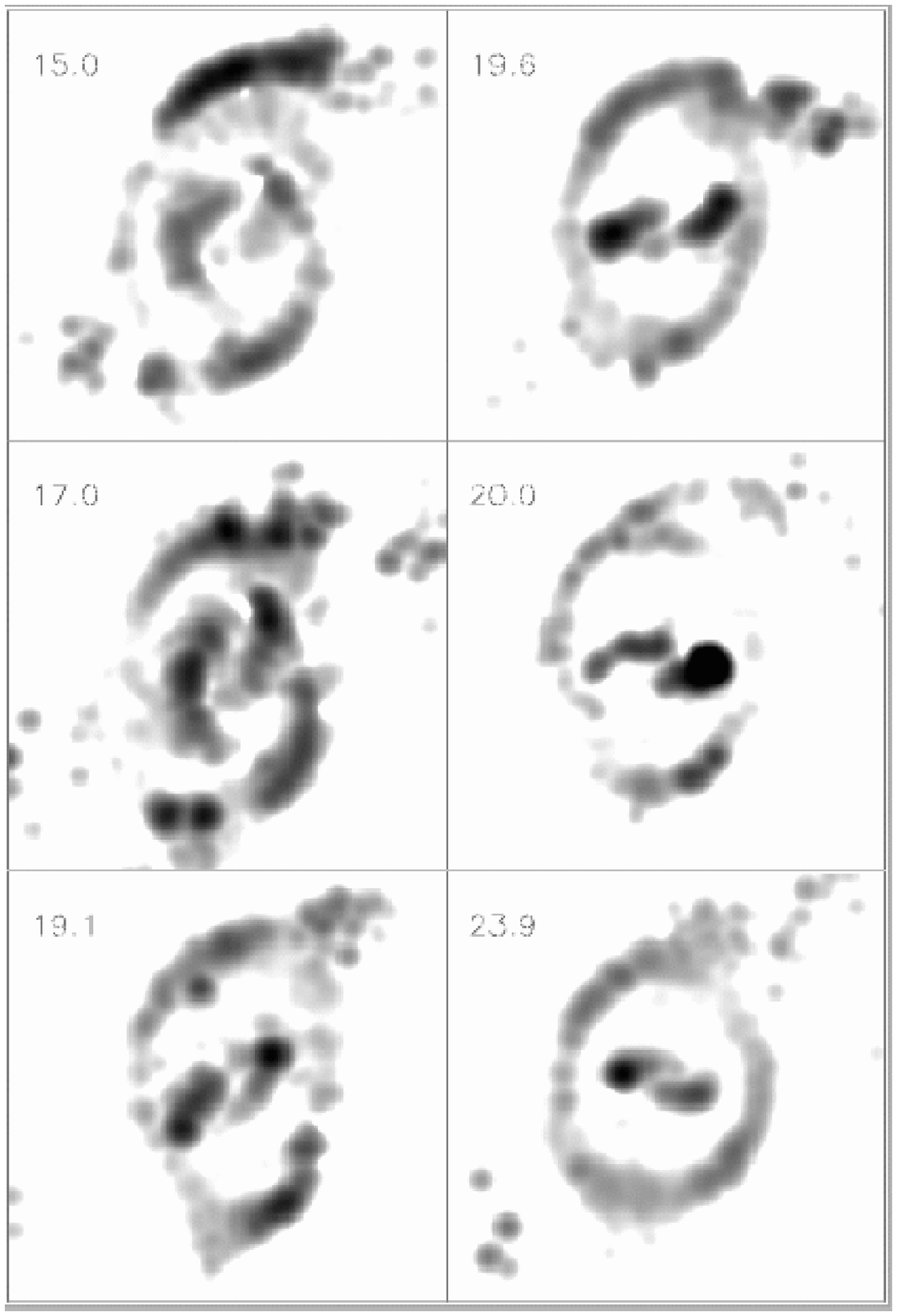}{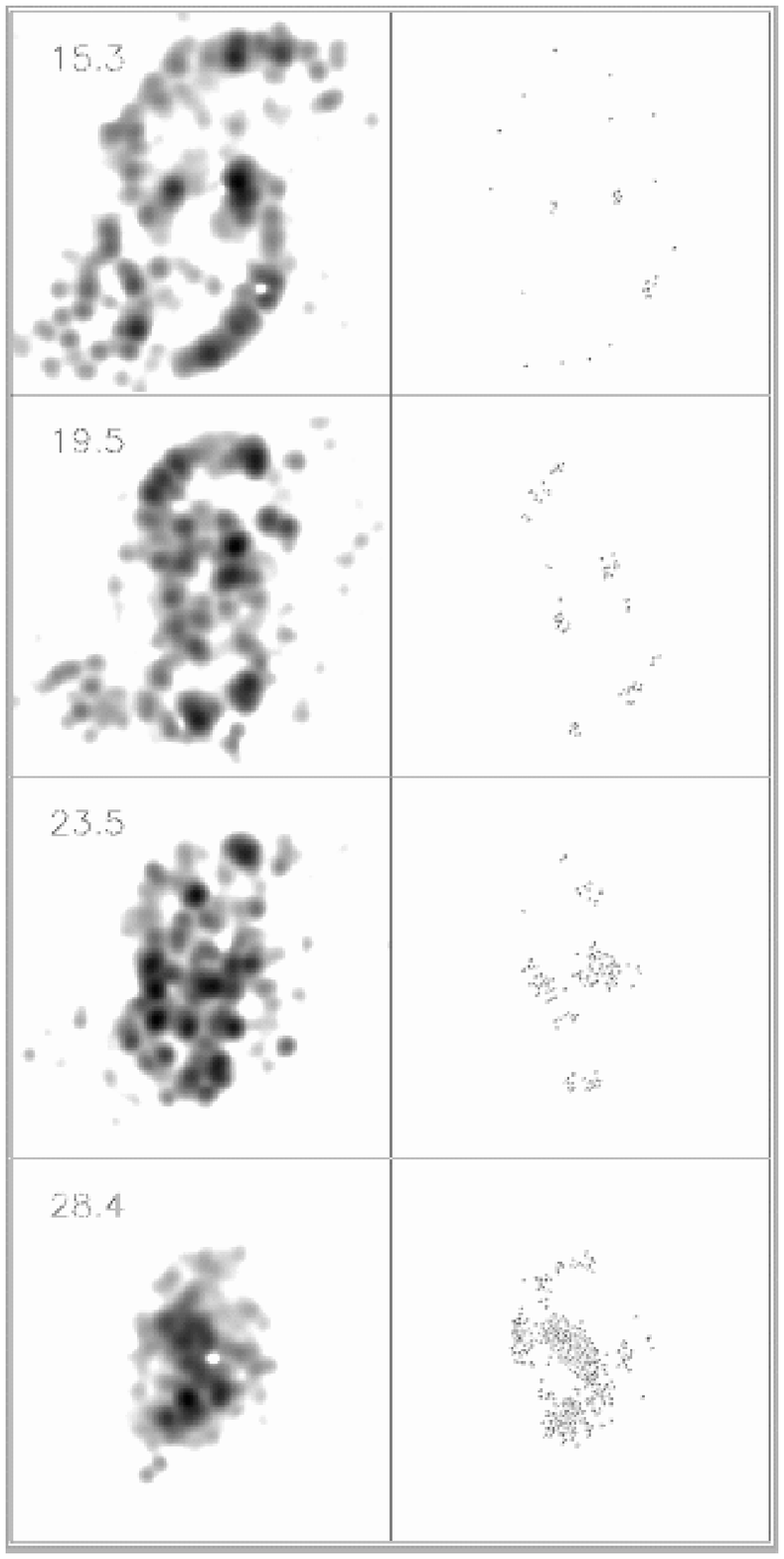}
\caption{Self-gravitating gas in self-consistent simulations of a 3D
gas $+$ stars disk (shown face on). Logarithmic {\it grayscale} map of shock
dissipation without ({\it 1st and 2nd columns\/}) and with 
({\it 3rd column\/}) star formation
in the central kpc of a barred galaxy with two ILRs. Time in bar rotations is
given in the {\it upper left  corners\/}. 
The stellar bar is horizontal and rotation is
anti-clockwise. The {\it right} ({\it 4th\/}) {\it column}
 is a star formation map corresponding
to the {\it third column\/}. 
Inclusion of star formation causes both nuclear rings to
merge and shows the subsequent gas inflow (last frame at 28.4) across the
inner ILR. ({\it From Knapen et al. 1995.})} 
\end{figure}  

The origin of the offset shocks in the presence of ILRs can be 
easily understood 
from the orbit analysis. With no ILRs, stationary periodic orbits, which
exist in the bar frame, are all aligned with the bar major axis (hereafter
$x_1$ orbits; Contopoulos \& Papayannopoulos 1980). Although the internal
dissipation in the gas leads to energy loss, one can still predict the major
features of gas response knowing the major families of periodic orbits in the
bar. At each ILR, the orbits change their alignment by $\pi/2$. Hence between
the ILRs (or between the center and the ILR, if only one exists) the orbits
are aligned with the minor axis of the bar (hereafter $x_2$ orbits). Stars can
change their orbits abruptly, while gas can do this only gradually by forming
a pair of shocks.

The most important observed characteristics of nuclear rings are: a) their
mass, typically in the range of $10^8$--a few $\times 10^9\ M_{\sun}$ in
 molecular gas; b) their single or double appearance; c) their
location close to the
turnover velocity; d) their oval or circular shapes leading the bar in the 
first
quadrant; and e) accompanying star formation in the inner part of the ring.
Theoretically, nuclear rings  are indeed subject to fragmentation and star
formation (Elmegreen 1994). This has been confirmed by numerical modeling, which
has shown also that the ring shapes, orientation, colors, and 
star formation distribution 
 depend on the rate of gas inflow from larger radii (Knapen et
al. 1995). Moderate or high rates of inflow result in a characteristic  oval
shape of the ring, with its major axis leading the bar by about
$50^\circ$--$80^\circ$, and with two or four prominent sites of star formation,
depending on
whether one or two ILRs exist (two star-forming regions on the bar
minor axis and two on the bar major axis, respectively, see Fig.~2). On the
other hand, low inflow rates result in much rounder and redder rings, with a
smoother distribution of star formation. In fact, no ring is found to be
sitting close to the outer ILR; rather, it is positioned half way between the
ILRs. If more than one ring is formed, rings can interact both
hydrodynamically and gravitationally (Heller, Shlosman, \& Englmaier 2001; see
also \S 4.1). 

Star formation increases the rate of viscous dissipation in the ring by
supporting turbulent motions in the gas. This appears to be one of
the factors capable of shrinking the ring radially and driving an inflow across
the inner ILR that is otherwise prevented due to the change in the direction
of the torque interior to the ring.    

Massive nuclear rings have been found to affect the stellar and gas dynamics,
in and out of the disk plane (Heller \& Shlosman 1996). This happens mainly by:
i) strengthening (or creating) double ILRs; ii) making the $x_1$
orbits interior to the ring less eccentric; iii) increasing the phase
space occupied by the $x_2$ orbits; and iv) skewing the $x_1$ orbits
outside an ovally shaped ring, misaligned with the bar major axis. This
leads to orbit intersection and depopulates them of the gas, which falls
towards the ring. 

Increasingly massive nuclear rings push the outer ILR outwards, weakening
the stellar bar between the ring and  corotation. When the outer ILR
radius becomes comparable to the bar minor axis, most of the orbits
supporting the bar potential disappear and the bar is expected to dissolve,
leaving a weakened bar remnant inside the inner ILR. Such bars, extending to the ILRs
are found in  late-type disk galaxies (Elmegreen \& Elmegreen 1985), with
no active star-forming rings encircling them. Bar dissolution outside the
nuclear ring will cut off gas inflow which fuels the star formation, although
aged stellar rings will survive dynamically (Shlosman 1999). Such red
``fossil'' rings have been found recently in a number of disk galaxies (Erwin,
Vega Beltr\'an, \& Beckman, this volume, p. 171).

To summarize, nuclear rings are clear signatures of large-scale stellar bars
and of the availability and recent radial transport of large quantitites of
molecular gas by means of gravitational torques.

\section{Nuclear Bars and Nuclear Spirals: Gas Dynamics}

\subsection{Gas Density Waves Within the Central kpc}

Two new morphological features can play a dominant role in the 
central regions of
disk galaxies. Recent high-resolution observations of galactic centers
revealed a number of paired, grand-design spirals (Laine et al. 1999; Martini
\& Pogge 1999; Regan \& Mulchaey 1999). Because most barred galaxies seem to
possess at least one ILR, and stellar density waves are damped and cannot
propagate across it, what drives the observed nuclear spiral structure? One
can be motivated by the fact that these spirals are notable in near-infrared
colors (of dust) and do not seem to be associated with a high level of star
formation. In addition, the inner spiral structure appears to be a continuation
of the outer spirals and shows low arm-interarm contrast. The above
constraints allow for an alternative explanation of grand-design two-arm
spiral structure in the centers of disk galaxies by means of gas density
waves (Englmaier \& Shlosman 2000). These can freely propagate across the
resonances. Probably the most interesting result is that the shape of the
nuclear spiral, i.e., its pitch angle, depends on the shape of the
gravitational potential and on the sound speed in the gas, hence providing an
insight into the mass distribution at several hundred~pc 
from the center and into
possible conditions in the ISM. The shape of the nuclear spiral
 requires a gently rising rotation curve
and hence a low central-mass concentration.

\begin{figure}[ht!!!!!!]
\vbox to3.0in{\rule{0pt}{3.0in}}
\includegraphics{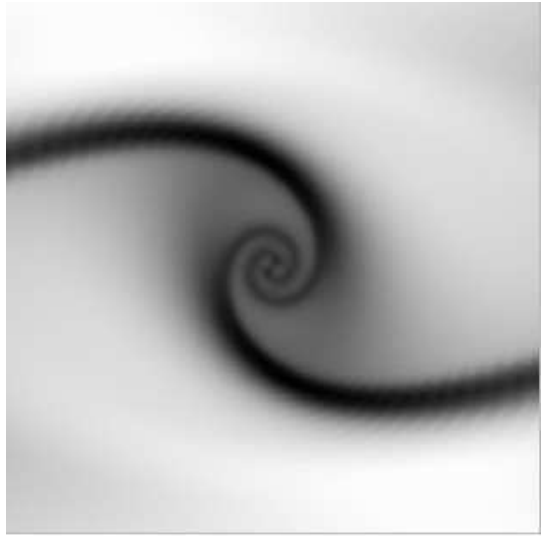}
\caption{{\it Grayscale} image of a steady-state gas response to the
bar perturbation in a model with mild central concentration (only the inner
2~kpc are shown). The stellar bar is horizontal and extends to 5~kpc and the
ILR is at 2 kpc; the gas rotates clockwise. The sound speed is 10~km~s$^{-1}$.
({\it From Englmaier \& Shlosman 2000.})} 
\end{figure}   
\begin{figure}[ht!!!!!!]
\vbox to2.3in{\rule{0pt}{2.3in}}
\includegraphics{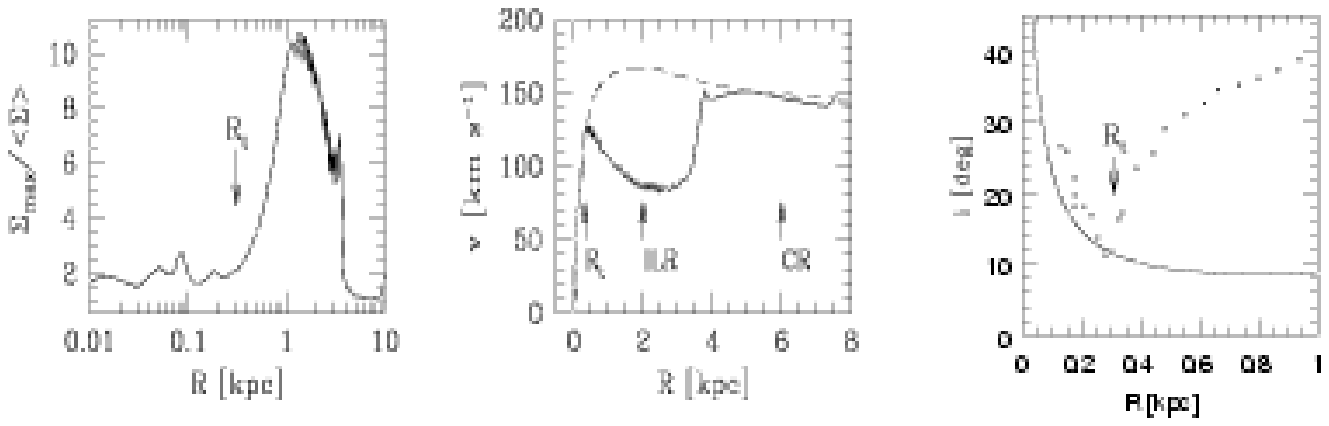}
\caption{{\it Left:\/} Spiral-arm surface density normalized by the
average density at radius $R$ for the standard model. The {\it 
vertical arrow} shows 
the position of a transition radius $R_{\rm t}$, where the spiral pattern winds up the
initial angle of $\pi$ (measured from corotation). The spiral weakens 
substantially inside $R_{\rm t}$ and outside  4 kpc where it crosses the bar
axis again. Shocks in the outer spiral ($>R_{\rm t}$) are unresolved,  thus the
surface density plotted here is only a lower limit.   
{\it Center:\/} The gas azimuthal velocity component ({\it solid line\/})
measured along the spiral in the inertial frame. Circular velocity for the
axisymmetric part of the potential ({\it dashed line\/}) 
is shown for comparison. The
gas is on nearly circular orbits inside $R_{\rm t}$ and outside 4 kpc, and
transits from $x_1$ to $x_2$ orbits in between.  
{\it Right:\/} Spiral pitch angle vs. $R$. Theoretical ({\it solid line\/}) and
modeled ({\it dots\/}) curves are shown. ({\it From Englmaier \& Shlosman 2000.)}}  
\end{figure}  

We specifically target the case where the large-scale perturbation
is caused by the stellar bar and therefore assume dominant barlike modes
with $m = 2$. This approach can be generalized for any symmetry of the
perturbation. For a certain  range of central concentrations and sound speeds,
$c_{\rm s}$, 
the gas response generates a spiral pattern between the corotation and
the center, winding into 
an unusual  angle of $\sim 3\pi$ radian (Fig.~3). Part of the
pattern corresponding to the winding of $\pi$ radians, i.e., between 
corotation and the first crossing of the bar major axis, is a strong standing
shock in the bar frame. We refer to the radius when the shock first crosses
the bar major axis as a transition radius, $R_{\rm t}$.  The spiral pattern is
formed by the gas moving from $x_1$ to $x_2$ orbits, which was confirmed by
means of non-linear orbit analysis (described in Heller \& Shlosman 1996).
The modeled gas response in Fig.~3, however, clearly shows  continuation
of the spiral pattern well past the region of expected winding.

To understand the reason for the observed gas response and the large winding
angle, we compare the density response amplitudes along the spirals
(Fig.~4, {\it left\/}). This amplitude decreases substantially after the first $\pi$,
counting from the corotation inwards, i.e., inside the $R_{\rm t}$. Such 
a decline
means that the shock strength drops sharply, and that
 the gas response at smaller
radii is more of a linear wave and is consistent with the gas settling on
the $x_2$ orbits. The initial $\pi$ winding corresponds to this process.  At
smaller radii, additional driving must support the modeled spiral pattern, a
process that requires a milder orbital change.        
                                 
The spiral pattern in Fig.~3 is stationary in the bar frame. Viewed in the
fluid rest frame at some $R$, the radial component of the wave vector points
inwards because the gas rotates faster than the spiral pattern everywhere
inside the corotation. It is convenient to explain the spiral in terms of a
wave packet propagating radially inwards, which is sheared by the
differential rotation in the disk.  The only waves which can propagate
inside the ILR are the short waves, and their group velocity is slower
than the sound speed, due to the contribution of the epicyclic motion.

The low arm--interarm contrast observed in the model inside $R_{\rm t}$
(Fig.~4, {\it left\/}) confirms that the wave is not 
accompanied by a strong compression.
This is in agreement with  observed nuclear spirals, which do not exhibit
extensive star formation and are mainly seen because of the dust obscuration,
and justifies our ignoring  self-gravitational effects in the gas. 
The pitch angle of the innermost spiral pattern decreases quickly with 
radius, i.e., the pattern becomes tightly wound. This is in sharp contrast
with the pitch angle at $R > R_{\rm t}$, which corresponds to the $x_1$-to-$x_2$
orbital switch. Here the pitch angle quickly increases with radius, and the
overall response becomes open. The wedge-shaped profile of the pitch angle in
Fig.~4 ({\it right\/}) is therefore characteristic of switching from one type of response to
another.

Both responses cannot coexist at the same radii.  Moving inwards from the
ILR, the spiral pattern associated with the outer shocks winds round the
center and the pitch angle decreases. At some value $R_{\rm t}$, the low-amplitude
density waves take over and slowly increase this angle. The gas can only
sustain waves of a particular frequency at each $R$; these
are solutions to the dispersion relationship. Therefore, the solutions for the
pitch angle can only switch from the inner to the outer one where they are
comparable. 

Importantly, there is a strong dependence of the spiral shape in the
nuclear region on the central concentration. A factor of 2 increase in central
concentration makes the spiral pattern so tightly wound that it is not
recognizable any more beyond the initial winding of $\pi$. This results in the 
formation of a featureless nuclear disk within the inner 400--500~pc. For
smaller central concentrations, the spiral becomes less tightly wound and
finally collapses towards the bar's major axis. The modeled gas responds in a
non-linear fashion, staying on $x_1$ orbits inside the weak ILR. We have
observed this effect, which leads to the formation of  shocks centered  on the 
major bar axis. Its strength depends also on the sound speed in the gas.
This effect has been studied independently by Englmaier \& Gerhard (1997) and
Patsis \& Athanassoula (2000). 
  
The measured pitch angle, $i$,
 of the modeled spiral is given by dots in Figure~4 ({\it right\/}).
We observe a twofold behavior for $i$; from the ILR inwards it is
decreasing until the shock pattern winds the angle of $\pi$ ($R_{\rm t}$, indicated
by an arrow), and then starts to unwind.  Remarkably, there is a sharp break in
the $i(R)$ slope at this point. The inner part of the
 $i(R)$ curve is in agreement
with theory. As we have noted above, the outer part of the $i(R)$ curve
describes the non-linear gas response to the bar forcing. Figure~4
 ({\it right\/}) demonstrates
that the large-scale shock penetrates inside the ILR and perturbs the nuclear
gas disk. Hence, it is the large-scale shock which directly drives the nuclear
spiral structure.            

Figure~4 ({\it middle\/}) displays the azimuthal velocity component 
of the gas flow measured
along the spiral arm. Inside the transition radius, $R_{\rm t}$, 
the azimuthal gas
velocity is very close to the rotational velocity in the axisymmetric
potential (dashed line). Somewhat farther out, the tangential velocity shows a
sharp drop, due to the non-circular motions in the gas which is in the process
of settling into $x_2$ orbits. This demonstrates the observational significance
of the transition radius, $R_{\rm t}$; namely, it represents the outer radius of the
nuclear disk where the gas is found in nearly circular $x_2$ orbits aligned
with the minor axis of the bar. The deviation from the axisymmetric rotation
curve is greatest around the ILR radius. Furthermore, the bar-driven shocks
reach the bar major axis again at $\sim 4\ {\rm kpc}$. Beyond this radius, the
gas follows nearly circular orbits.   

An alternative way to explain  nuclear spirals exists, at least in
principle, and requires a strong central-mass concentration in the form,
for example, 
of a supermassive black hole (SBH). One might expect that the SBH not only
contributes its own mass but  
affects the nearby stellar density distribution
as well---a kind of gravitational ``polarization''  effect which encompasses a
factor of ten more mass than the SBH. This will lead to the formation of an
additional (NLR) resonance and a pair of trailing spirals (Shlosman 1999) that
can have an appearance of grand-design nuclear arms. 

In the first case, the nuclear spirals are not expected to have a strong
effect on the radial gas redistribution due to their low amplitude. This
may not be true in the latter case, as the shocks will quickly move the gas
across the NLR.

\subsection{Nested Bars}

A morphological feature that is expected to dramatically affect the
gas dynamics inside the central kpc is a nuclear bar. Although these sub-kpc
bars have been probably detected by de Vaucouleurs (1974), Sandage \&
Brucato (1979), and  Kormendy (1982) as optical isophote twists in the central
regions of barred galaxies, they have been interpreted as triaxial bulges.  
High resolution ground-based observations have revealed a number of galaxies
with small bars residing within the large-scale stellar bars (e.g., Buta \&
Crocker 1993; Shaw et al. 1995; Friedli et al. 1996; Jungwiert, Combes, \& Axon
1997; Mulchaey \& Regan 1997; Elmegreen, Chromey, \& Santos 1998; Jogee, Kenney, \& Smith
1998; Erwin \& Sparke 1999, Knapen, Shlosman, \& Peletier 2000; Emsellem et al.
2001; Laine et al. 2001). The molecular gas content of these bars varies. In
some cases, the cold gas can be dynamically important, as evident from the 
 2.6~mm CO emission and NIR lines of H$_2$ (e.g., Ishizuki et
al. 1990; Devereux, Kenney \& Young 1992; Kotilainen et al. 2000; Maiolino et
al. 2000). Depending on the gas fraction contributing to the gravitational
potential, one can distinguish  stellar- or gas-dominated and
mixed nuclear bars. The largest sample (112) of disk galaxies analyzed so far
for this purpose reveals a substantial fraction of double (nested) bars,
probably in excess of 20--25\%, and, in addition, that
about 1/3 of barred galaxies
host nested bars (Laine et al. 2001).  

Probably the most intriguing property of nested bars is their theoretically
anticipated stage of a  dynamical decoupling, when each bar exhibits a
different pattern speed (Shlosman, Frank, \& Begelman 1989). Several aspects
of this problem, which are closely
 related to the formation of double-bar systems and dynamical
runaways there, are analyzed in the next section. Here we
focus on the important characteristics of gas flow in double bars,
presumed to tumble with pattern speeds $\Omega_{\rm s} > \Omega_{\rm p}$,
where ``p'' stands for the primary (large-scale) bar and ``s'' for the
secondary (sub-kpc) bar.

The gas response in  nested bar systems has been never studied in detail.
The first numerical simulations of double bars (Friedli \& Martinet 1993;
Heller \& Shlosman 1994)
were aimed at showing that bars can decouple. Elsewhere, in observational studies
of nuclear bars, it was assumed that the
 properties of nuclear bars are identical
to those of large-scale bars, in particular their gas dynamics and
the appearance
of characteristic offset dust lanes (e.g., Regan \& Mulchaey 1999; Martini \&
Pogge 1999). But numerical simulations and general theoretical
considerations support the view that nuclear bars are not scaled-down
versions of large bars, and the gas response in  nested bar systems differs
profoundly from that in single bars (Shlosman \& Heller 2001). This difference
results from the underlying gravitational potential being time-dependent in
all frames of reference, when  bars tumble
with different pattern speeds, and a number of other factors. In such  cases
the Jacobi energy (e.g., Binney \& Tremaine 1987) is not an integral of motion
even for a dissipationless ``fluid'', i.e., stars.

To understand the gas flow in  time-dependent nested-bar potentials,
one needs to study the dynamical constraints operating in these systems,
such as conditions minimizing  chaos. Pfenniger \& Norman (1990) have
pointed out that, in order to decrease the fraction of chaotic orbits in the
system, corotation of the secondary bar must lie in the vicinity of the
primary-bar ILR. This constrains $\Omega_{\rm s}$. 
Such a dynamical configuration, in
principle, poses a problem for uninterrupted gas inflow towards smaller
radii. Namely, the gas flow is repelled at the bar corotation, inwards or
outwards, because of the rim formed by the effective potential there, and
hence may not cross the ILR/corotation radius (i.e., the bar--bar interface).
Shlosman et al. (1989) have argued, in essence, that it is the gas
self-gravity that overcomes such repulsion by modifying the underlying
potential. In fact, even in the limit of neglibigle self-gravity in the gas,
the flow is capable of crossing the bar--bar interface, but not in a steady
manner and only for a restricted range of azimuthal angles (Shlosman \& Heller
2001). Independently, one can analyze the conditions for shock and dust-lane
formation in nuclear bars using a cloud--cloud collision model for the ISM and
show that cloud properties and number densities should be extreme in order to
maintain the cloud mean free path well below the nuclear bar width. 

\subsubsection{Modeling the Gas Response in Nested Bars}

Numerically, one can best study the gas response to the nested-bar potential  
by defining three regions in the disk: i) the primary-bar region
(outside its ILR), ii) the bar--bar interface (hereafter the {\it
interface}) encompassing the outer ILR of the large bar and the outer part of
the secondary bar, and iii) the interior of the secondary bar.   
In the first region, the primary bar outside the interface, the gas responds
by forming a pair of large-scale shocks, corresponding to the offset dust lanes,
and flows inwards across the interface into the secondary bar.  The flow in 
the primary bar, outside the interface, is steady and the shock strength and
shape are nearly independent of time. Our analysis is limited to this time
only and avoids transients.   

\begin{figure}[ht!!!!!!]
\vbox to2.5in{\rule{0pt}{2.5in}}
\includegraphics{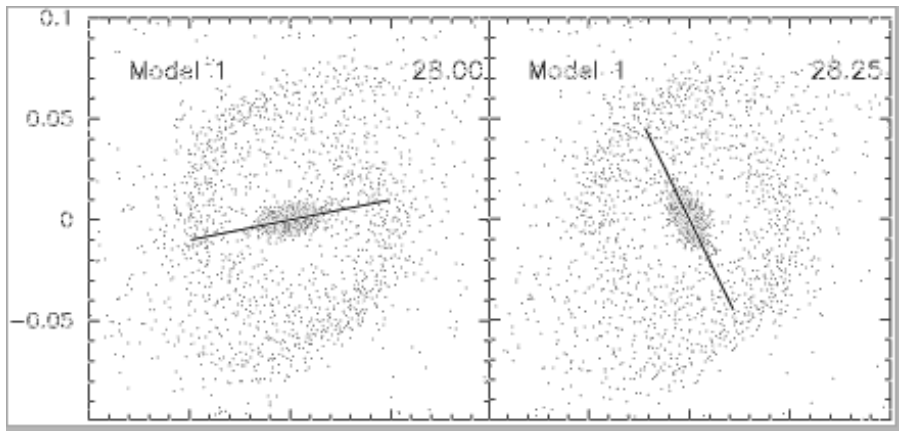}   
\caption{Snapshots of the gas density distribution in the central
kpc ($=0.1$): 2D SPH simulation in the background gravitational
potential of a nested-bar disk galaxy (face on). The gas response to the
bar torquing is shown in the primary bar (horizontal) frame of
reference. Both bars and gas rotate counter-clockwise. The position angle of
the secondary bar and its length are indicated by a {\it straight line\/}. 
Note
that the nuclear bar lies within the nuclear ring. Time is given in units of dynamical
time. ({\it From Shlosman \& Heller 2001.})} 
\end{figure}  
\begin{figure}[ht!!!!!!]
\vbox to3.6in{\rule{0pt}{3.6in}}
\includegraphics{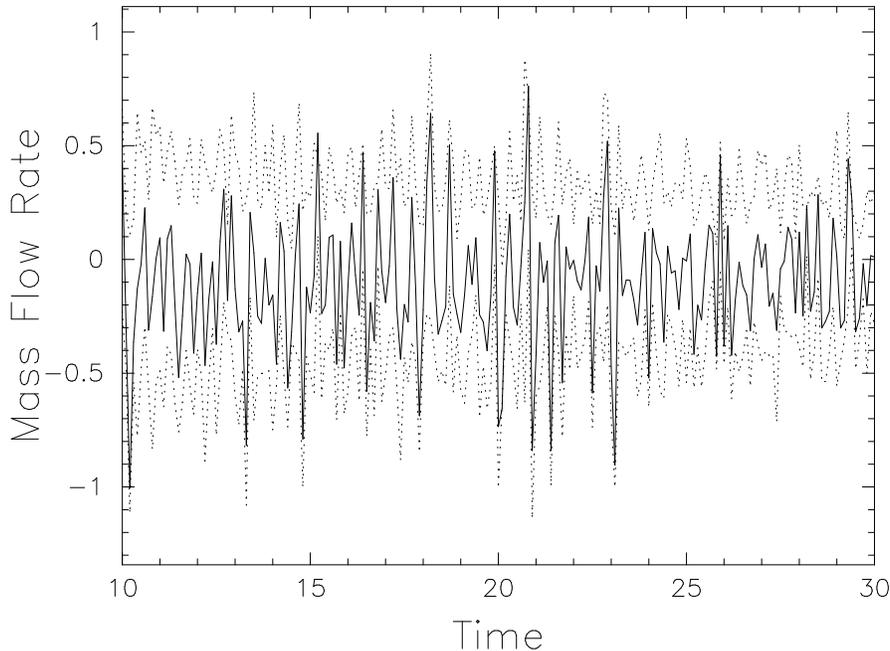}   
\caption{Time evolution of the gas inflow (negative) rate across the bar--bar
interface of a double-bar system. {\it Upper dotted line:\/} Flow rate within
$\pm 45^\circ$ of the major axis of the primary bar. {\it Lower dotted line:\/}
Flow within $\pm 45^\circ$ of the minor axis of the primary bar. {\it Solid
line:\/} Total flow across the bar--bar interface. Time is given in units of
dynamical time.({\it From Shlosman \& Heller 2001.})} 
\end{figure}  

In the second region, the bar--bar interface, the flow is time-dependent due
to the perturbative effects of the secondary bar and changing background
potential (Fig.~5). It correlates with the position angle of the secondary bar with
respect to the primary bar or can be erratic. We have subdivided the azimuthal
dependence of the gas flow into $\pm 45^\circ$ with the major axis of the
primary bar and $\pm 45^\circ$ with its minor axis and Fourier-analyzed it. 
  We
found the beat frequency of the secondary bar for all
models. But for smaller bars (i.e., short of their corotation) the
amplitude of this Fourier component is lower and accompanied by other frequencies.
For large nuclear bars (i.e., extending to their corotation), having a substantially
stronger influence at the interface,  the inflow shows the beat frequency
clearly identified with the secondary bar tumbling (Fig.~6).
 {\sl So the flow
across the bar--bar interface depends upon the strength of the secondary bar},
ranging from chaotic, for a relatively weak perturbation of the secondary
bar, to a more regular one. The corresponding mass
influx rate is of the order of $0.3\ M_{\rm gas,9}\ M_\odot$ yr$^{-1}$,
where $M_{\rm gas,9}$ is the total gas mass in the disk in units of
$10^9\ M_\odot$ within the corotation of the large bar. On  average, the
inflow proceeds through the broad region along the primary-bar minor axis,
while an outflow (albeit at a smaller rate) is directed along its major axis.
The reason for this behavior is that the inflow is driven mainly along the
large-scale shocks penetrating the bar--bar interface from the primary bar. At
the same time the outflow is detected at angles which do not encompass the
large-scale shocks. The net effect is clearly an inflow across the corotation
of the secondary bar.                 

The gas which is repelled by the secondary bar along the major axis
of the primary bar is found to enter large-scale shocks while still moving
out. This increases the mixing of material with  different angular momentum
increases inflow along the shocks. After crossing the bar interface the gas 
does not settle down but falls towards the third region, the inner 1/2--1/3
of the bar. Within this region the flow is very relaxed  at all times, with
uniform dissipation (well below the maximum dissipation in the large-scale
shocks), and no evidence for grand-design shocks. In fact, we observe a
kind of a ``limb brightening'' at the edge of this bar. This is seen in
Fig.~7  as an enhanced density of above-average dissipating particles
outside the oval-shaped central region and happens because the gas
joins the bar from all azimuths.     

\begin{figure}[ht!!!!!!!!!!!!!!!!!!!!!!!!!]
\vbox to5.9in{\rule{0pt}{5.9in}}
\includegraphics{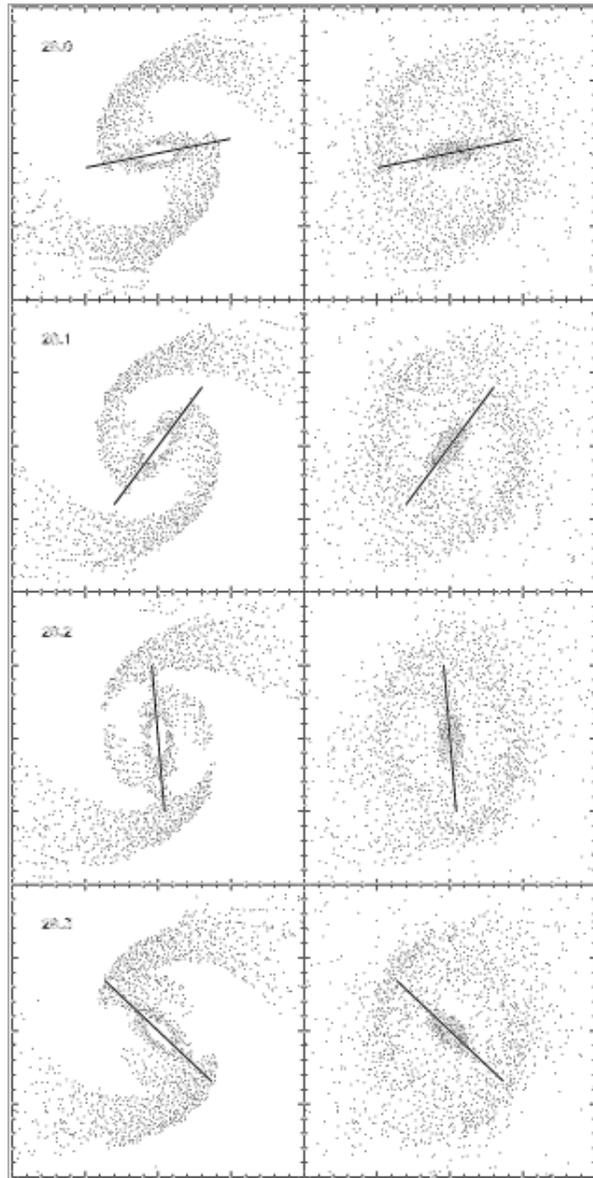}
\caption{Pattern of shock dissipation ({\it left\/}) 
and density evolution ({\it right\/}) in
the central kpc, shown in the frame of reference of the
primary bar (horizontal). Positions of the secondary bar and its length are
indicated by a {\it  straight line\/}. All rotation is counter-clockwise. The particles
on the {\it left}
 are those having a greater-than-average dissipation rate, which is
given by the time derivative of the non-adiabatic component of internal energy.
Note the sharply reduced dissipation in the innermost secondary bar and 
the ``limb
brightening'' enveloping it. Also visible are two dissipative systems associated
with the large-scale shocks in the primary bar and with the trailing shocks in
the secondary bar. ({\it From Shlosman \& Heller 2001.})} 
\end{figure}       

The pattern of shock dissipation in nested bars can be inferred from Figure~7.
It allows one to separate the incoming large-scale shocks from those driven
by the secondary bar. Note that two systems of spiral shocks occur in Fig.~7,
each associated with its corresponding bar. The large-scale (hereafter
``primary'') shocks extend to just pass off the minor axis of the primary bar,
when the secondary bar is nearly orthogonal to the primary one. As it
continues to tumble, the primary shocks extend deeper into the small bar.
Sometime after both bars are aligned, the outer shocks detach themselves
 from the small
bar, which is left with an additional pair of trailing shocks. This effect is
especially pronounced between $t=28.1$ and 28.2. To summarize,
the interaction between these shock systems shows detachment when the bars are
perpendicular and attachment when they are aligned with each other. The shapes
of the shocks depend on the angle between the bars.

\subsubsection{The Absence of Offset Shocks in Nuclear Bars}

A number of factors differentiate the gas dynamics in the decoupled secondary
bars from that in the single bars. The first is the time-dependent nature of the
gravitational potential in the nested-bar system during the dynamically 
decoupled 
phase. The second is the gas injection into the secondary bar which proceeds
through the primary large-scale shocks penetrating the bar--bar interface. 
Such a
phenomenon is absent at the corotation of the primary bars, which  is
depopulated of gas. A large amount of dissipation accompanying this
process results in an inability of the gas to settle in the outer half of the
secondary bar, which raises the question of whether the secondary bars can
extend to their corotation.

The third factor 
is a fast secondary-bar pattern speed that does not permit 
secondary ILRs to form. If the decoupled phase of nested bars is
short-lived, the quadrupole interaction between the bars will not be able to
brake the small bar and form the ILRs. However, even in the case of a
long-lived decoupled phase we do not expect  nuclear bars to slow down, as
the gas inflow across the interface and the resulting central concentration in
fact speeds a bar up, as was found by  Heller, Noguchi, \& Shlosman (1993,
unpublished). Because of the fast rotation, the $x_1$ orbits deep inside
the secondary bars are round, with no loops or needle-shapes. The 
low-Mach-number gas flow is well organized here and capable of following these
orbits with little dissipation, as shown in Figure~7. Non-linear orbit
analysis reveals that the main orbits aligned with the secondary bar, $x_1$,
have a mild ellipticity and no end-loops. This result is quite robust and
holds despite the extreme axial ratio, $4:1$, used here. No offset large-scale
shocks form under these conditions.

Decoupled secondary bars avoid their ILRs because of their high pattern
speeds. When  ILRs are absent in a large-scale bar, the offset shocks
weaken and recede to the major axis of the bar, becoming ``centered'' (e.g.,
Athanassoula 1992). They do not disappear completely because the underlying
stellar periodic orbits have  end-loops, are pointed, or have large
curvature at the ends, forcing the gas to shock there. These orbital shapes
result from the slower rotation of large,  compared to nuclear, bars.
Despite the fact that such weak centered shocks can exist theoretically, only
two examples have been found  from more than a hundred barred galaxies
analyzed by Athanassoula (1992). During the last decade only one more
potential example has been added to this list (Athanassoula, private
communication). It seems plausible that centered shocks are very rarely
observed because they are so weak.

Instead of ``classical''
 centered shocks we find that the {\sl inner} half of the
secondary bars shows a rather uniform dissipation during the early stages of
the gas inflow, which sharply decreases with time, or a ``limb brightening''
effect (Fig.~7). This dissipation
is always small compared to dissipation in the primary large-scale shocks.
Knapen et al. (1995) have analyzed the shock  dissipation in a self-consistent
gravitational potential of ``live'' stars and gas {\sl before} the onset of
decoupling, when both bars tumble with the same pattern speeds, and when the
gas self-gravity is accounted for. No offset shocks have been found in this
configuration either.

An important issue is the fate of the gas accumulating in the inner parts of
the secondary bars. Under the observed conditions in numerical
simulations (gas masses and surface densities) the gas self-gravity should
exert a dominating effect on its evolution (see \S 4.2). 

We conclude that no large-scale shocks and consequently no offset dust lanes
will form inside secondary nuclear bars either when they are dynamically coupled
and spin with the same pattern speeds as the primary bars, or when they
are dynamically decoupled and
spinning much faster. Two main factors, the time-dependent gravitational
potential and the nature of the gas flow deep inside the bar, prevent the
formation of these dust lanes.  The time-dependent, ``anisotropic''  gas
inflow across the ILR/corotation interface found here is a completely new
phenomenon inherent to nested bars, because in  single-barred systems no
such inflow is possible at all {\sl  when the gas accumulation is small and
its self-gravity is negligible\/}. The fate of the gas settling inside 
nuclear bars cannot be decided without invoking global self-gravitational
effects in the gas that will completely change the nature of the flow there.

\section{The Formation and Dynamical Decoupling of Nuclear Bars}

Despite the obvious importance of the nested-bar configuration, our
understanding of its formation, dynamics, and evolution is very limited.
Even the sense of rotation of nuclear bars and their pattern speeds have
never been detected directly from observations, although a number of projects
are  working on this.

Theoretically, three options exist. First, if the nuclear bar was formed via
self-gravitational instability (in stellar or gaseous disks), it must spin
in the direction given by the angular momentum in the disk, i.e., in the
direction of the primary bar (Shlosman et al. 1989). In this
case, the pattern speed of the nuclear bar, $\Omega_{\rm s}$, must be substantially
{\sl higher} than that of the primary bar, $\Omega_{\rm p}$,
 which has been confirmed in
numerical simulations (Friedli \& Martinet 1993; Combes 1994; Heller \&
Shlosman 1994). The  presence of gas appears to be imperative for this to occur 
(e.g., Shlosman 1999). Both bars are dynamically {\sl decoupled} and the
angle between them in a face-on disk is arbitrary.         

Second, two bars can corotate, being dynamically {\sl coupled} and their
rotation completely synchronized. Such a configuration can be a precursor
to the future decoupled phase (discussed above), or continue indefinitely
(e.g., the numerical simulations of Shaw et al. 1993; Knapen et al. 1995).
The nuclear bar is expected to lead the primary bar in the first quadrant
at a constant angle, close to $90^\circ$. An explanation for this phenomenon
lies in the existence of two main families of periodic orbits,
$x_1$ and $x_2$, in barred
galaxies (see \S 2). The gas, losing angular support
due to the gravitational torques from the primary bar, will flow towards the
center and encounter the region with the $x_2$ orbits, which it will
populate\footnote{Strictly speaking, the gas will not occupy  perfectly
periodic orbits because of dissipation, but will be found at nearby energies,
which is sufficient for our discussion.}. The extent of the $x_2$ family
defines the inner resonance region in the disk, i.e., the position of the
non-linear inner and outer ILRs. The nuclear bar may be further strengthened by
the gas gravity, which drags stars into $x_2$ orbits. However, the amount of gas
accumulating in the ILR resonance region may be insufficient to cause the
dynamical runaway. In this latter case, it is thought that the synchronized
system of two nearly perpendicular bars can be sustained indefinitely, until
star formation or other processes act.

Third, the secondary bar can rotate in the opposite sense to the primary bar.
This situation may arise from merging, when the outer galactic disk acquires
opposite angular momentum to that of the inner disk. This was considered by
Sellwood \& Merritt (1994) and Davies \& Hunter (1997) and appeared as a
non-recurrent configuration. At least one of the bars should be purely
stellar, because  the gas cannot populate intersecting orbits.               

Although all three options discussed above make specific predictions verifiable
observationally, the triggering mechanism(s) for the formation of such
systems require much better understanding. Much of the computational
effort has so far gone into studying decoupling in self-gravitating (gas and
stars) systems (e.g., Friedli 1999; Shlosman 1999). Unexpectedly, even
non-self-gravitating gas can form a bar and decouple from the background
potential (Heller et al. 2001). Following this work, we show
that dynamical evolution does not stop with the formation of two coupled
perpendicular bars, {\sl even} when gas self-gravity is neglected. Instead,
partial or complete decoupling of a nuclear gaseous bar, depending on the
degree of viscosity in the gas, is triggered for a prolonged  period of time.
This bar either librates around the major axis of the primary bar, or acquires
a different pattern speed, which is substantially {\sl slower} than that of
the primary bar. The role of gas self-gravity in the decoupling
process is discussed afterwards, and in this case $\Omega_{\rm s} > 
\Omega_{\rm p}$. Both
regimes are likely to be encountered in nature and are therefore relevant to our
understanding of galactic evolution.   

\subsection{Non-Self-Gravitating Nuclear Gaseous Bars}  

\begin{figure}[ht!!!!!!!!!!!!!!!!!]
\vbox to6.3in{\rule{0pt}{6.3in}}
\includegraphics{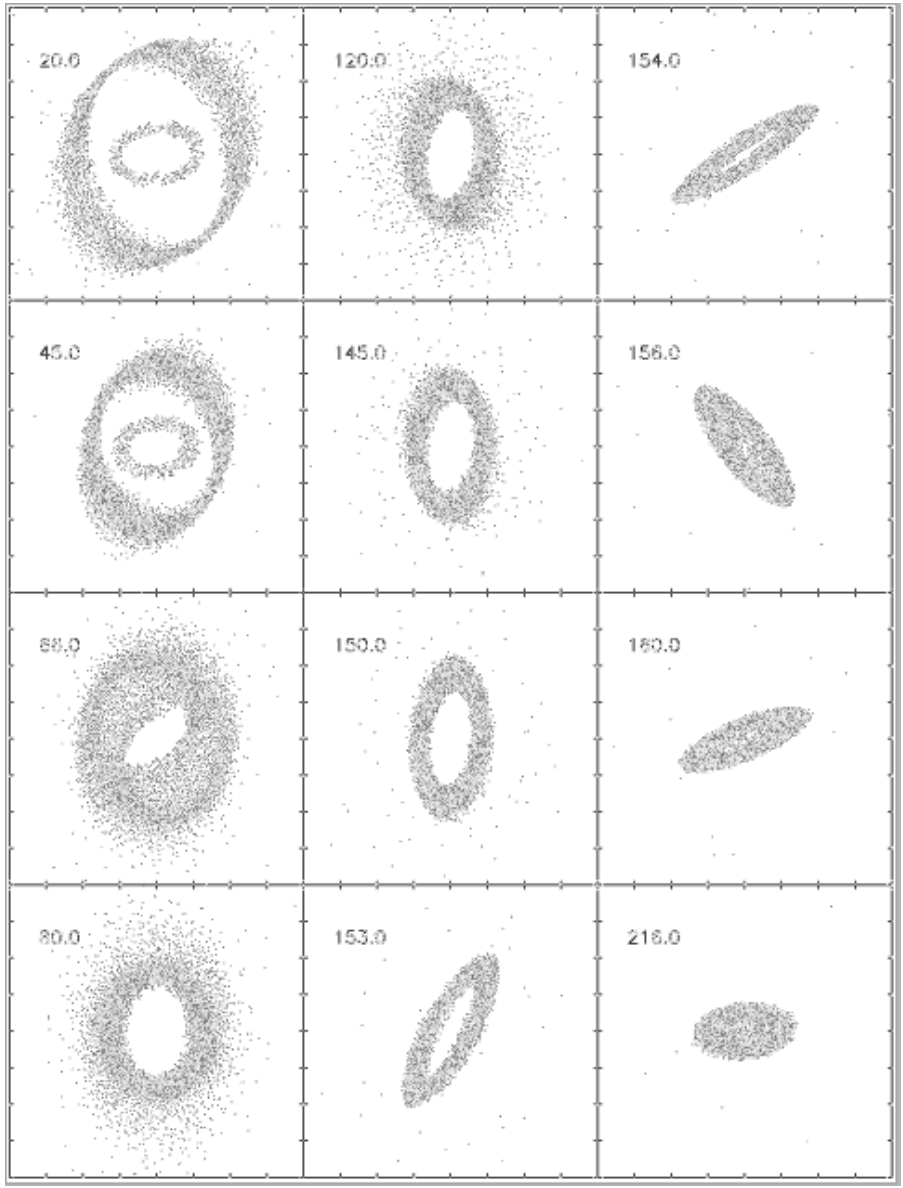}
\caption{Time evolution of the low-viscosity model: 2D SPH simulation in the
background gravitational potential of a barred disk galaxy (shown face on).
The gas response to the bar torquing is displayed in the primary-bar frame.
The primary bar is horizontal and the gas rotation is counter-clockwise. Note
a fast evolution after $t\sim 150$, when the nuclear bar decouples and swings
clockwise (!). The bar is ``captured'' again at $t\sim 211$. Time is given in
units of dynamical time, $\tau_{\rm dyn}$. The animation sequence discussed
 here 
and others are available in the online edition of Heller et al. (2001).} 
\end{figure}           

It is a common wisdom that a low-surface-density gas responds to  single-bar
torquing by forming a pair of offset shocks driving the gas towards the
nuclear ring(s). Any further evolution stops here. Unexpectedly, we find
that even non-self-gravitating gas can form a (gaseous) bar and decouple from
the background potential, depending on the viscosity of the ISM (Heller
et al. 2001). 

The actual degree of viscosity in the ISM is difficult to estimate, partially
due to its multiphase character. We have run three numerical models with identical
initial conditions, with the sole difference that the value of the viscosity was 
changed by a factor of 2 above and below that of this standard model (hereafter
``high'' and ``low'' viscosity models). The gas was evolved using a 2D version of
an SPH code (details in Heller \& Shlosman 1994), neglecting the gas
self-gravity and, alternatively, using the grid code ZEUS-2D (Stone \& Norman
1992). Both numerical schemes gave similar results.  

The most spectacular evolution  occurred in the low-viscosity model, although
all the models showed similar initial evolution during which, after forming a pair
of large-scale shocks, the gas lost its angular momentum and accumulated in a
double ring, corresponding roughly to the inner and outer ILRs. 
Furthermore, in all the
 models the rings interact hydrodynamically, destroying one
of them. In the low viscosity model (Fig.~8), the inner ring is destroyed
and its gas is mixed with the outer ring at intermediate radii. After
merging, a single oval-shaped ring corotates with the primary bar, leading it
by $\phi_{\rm dec}\sim 50^\circ$ (the so-called decoupling angle, whose value 
depends on the gas viscosity).  

\begin{figure}[ht!!!!]
\vbox to2.4in{\rule{0pt}{2.4in}}
\includegraphics{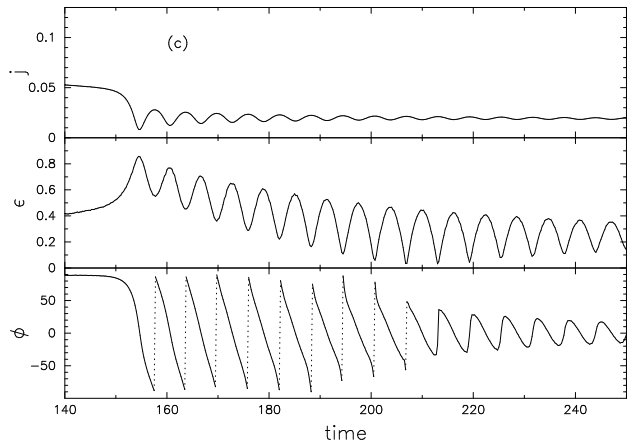}
\caption{Low-viscosity model (see also Fig.~8): evolution of total specific angular 
momentum $j$ in the primary-bar frame of reference ({\it upper\/}), eccentricity
$\epsilon = 1-b/a$ ({\it middle\/}), and position angle, $\phi$, of the line of apsides
({\it lower\/}) of the gaseous nuclear bar. Both $\phi$ and $\epsilon$ were computed
from the tensor of inertia of the gas within the small bar. The time is given
 in
units of dynamical time.
The angle $\phi$ is shown between $\pm\pi/2$ for
presentation only and discontinuities are marked by {\it vertical dotted lines\/}.
Full decoupling of the nuclear bar exists at $\tau\approx $ 150--211.}
\end{figure}      

Subsequently, the remaining ring (in all the
 models) becomes increasingly oval and
barlike (see eccentricity evolution for low viscosity model in Fig.~9), its
pattern speed changes abruptly, and it swings towards the primary bar, {\sl
against} the direction of rotation of this bar. We describe the evolution of
this gas flow in terms of the bar dynamics and call it a {\it nuclear gaseous
bar}. (In the inertial frame, the nuclear bar still spins in the same direction
as the main bar, albeit with a smaller pattern speed than $\Omega_{\rm p}$.) The
shape of this bar can be described by its axial (minor-to-major) ratio, which
reaches a minimum when both bars are aligned. After crossing the primary-bar
axis into the fourth quadrant, the nuclear bar initially decreases its axial
ratio, i.e., it becomes less oval. Thereafter it stops, becomes more oval, and
speeds up in the prograde direction. The bar axial ratio reaches its
minimum when both bars are aligned and the gaseous bar rotates in the
retrograde direction (in the primary-bar frame!). 

In standard and high-viscosity models, the nuclear bar librates about
the primary bar with a decreasing amplitude, being damped more strongly in the
latter model. The low-viscosity model, however, behaves in a
qualitatively different way. Instead of librating about the primary bar, the
nuclear bar continues to swing in the same direction maintaining a pattern
speed $\Omega_{\rm s} < \Omega_{\rm p}$ for about 60 dynamical times, corresponding in our
units to about 2--3 $\times 10^9$~yr. On average, its pattern
speed is about half of $\Omega_{\rm p}$, oscillating around this value with a
substantial amplitude. At times, $\Omega_{\rm s}$ is very small, giving the
impression that the nuclear bar stagnates in the inertial frame of reference.
With time, the eccentricity of the nuclear bar gradually decreases and the bar
is trapped again by the valley of the potential of the main bar, thus entering
the libration phase, similar to other models.           

What are the physical reasons for the partial or complete runaways shown
in each model? Looking at the distribution of gas particles in the rings with
Jacobi energy, $E_{\rm J}$, one finds that after merging the ring is positioned
close to the energy where the transition from $x_2$ to $x_1$ (at the inner ILR)
occurs. The exact value of this transition energy is model-dependent of course,
but this is of no importance to the essence of the decoupling.          

Viscosity is most important during the phase when both rings interact
hydrodynamically and merge, which understandably starts earlier and proceeds
faster for higher-viscosity models. This pre-decoupling evolution is similar
in all models, apart from the timescale. The crucial difference between the
models comes from 1) the position of the nuclear bar on the $E_{\rm J}$
axis after 
merging and 2) the value of the $\phi_{\rm dec}$ angle, i.e., the orientation of
the nuclear bar at the time of decoupling. We note that the role of viscosity
here is fundamentally different from that in nearly axisymmetric potentials
(planetary rings, galactic warps, etc.), where it acts to circularize the orbits. In
barred potentials, circular orbits do not exist and the gas merely moves from
one elongated orbit to another.

The single remaining ring is subject to a gravitational torque from the
primary bar, which acts to align the ring with the bar. In the pre-decoupling
stage, the ring becomes progressively barlike, decreasing its axial ratio. In
the low-viscosity model, this gaseous bar resides in purely $x_2$ orbits and,
therefore, responds to the primary-bar torque by speeding up its precession
while being pulled backwards until it is almost at  right angles to the bar
potential valley. The decoupling happens  abruptly when a substantial
fraction, $\sim1/2$, of the gas in the bar finds itself at $E_{\rm J}$ energies
below the inner ILR. The absence of $x_2$ orbits at these $E_{\rm J}$ 
means that the
bars either are unable to settle on these orbits or lose their stable
orientation along the primary-bar minor axis. As a result, the nuclear bars
enter into libration about the main bar with smaller or larger initial departures
from the stable orientation along the main bar, {\sl which is a single
decisive factor separating the partial and full decouplings\/}.

The angle $\phi_{\rm dec}$ is largest for the low-viscosity model---the reason
for the qualitatively different behavior. All three models show that 
gaseous bars in the partial or full decoupling phase experience shape changes
depending on their orientation. The bar has a much smaller eccentricity in the
fourth quadrant than in the first  (Fig.~9). Such an asymmetry with respect
to the primary-bar axis ensures that the resulting gravitational torques from
the primary bar are smaller in the fourth quadrant. But only for the least
viscous model, which has the maximal $\phi_{\rm dec}$, does this
make the ultimate difference. In this case the torques are unable to confine
the bar oscillation, which continues for a full swing of $2\pi$. The nuclear
bar is trapped again at $\tau\sim 211$, after many rotations with respect to
the large-scale bar.  

The clear correlation between the eccentricity of the nuclear bar and
its angle with the primary bar in the decoupled phase shown by {\sl all}
the 
models can be tested observationally, e.g., by looking at the molecular rings
in CO. The maximum  in the gaseous bar eccentricity, $\epsilon$,
 is achieved each
time both bars are aligned and the minimum occurs when the bars are at right
angles.

Two additional effects should have observational consequences as a
corollary to decoupling and the periodic increase in $\epsilon$. First,
the gas will move inwards across the inner ILR on a
dynamical timescale. Shlosman et al. (1989) pointed out that  ILR(s)
present a problem for radial gas inflow because the gas can stagnate there. A
solution was suggested in the form of a global self-gravitational instability
in the nuclear ring or disk, which will generate gravitational torques in the
gas, driving it towards the nuclear region.  

Recently  Sellwood \& Moore (1999) resurrected the idea that ILRs would
``choke'' the gas inflow. However, as we see here, even non-self-gravitating
nuclear rings are prone to dynamical instability which drives the gas inwards,
whether in a full or partial decoupling. Second, this instability can be
accompanied by star formation along the molecular bar due to increased
dissipation in the gas, and this star formation will have a quasi-periodic
bursting character.

\subsection{Self-Gravitating Nuclear Gaseous Bars} 

Gas self-gravity plays a crucial role in the formation and dynamical decoupling
of secondary, nuclear, bars. In this case $\Omega_{\rm s} > \Omega_{\rm p}$
(Shlosman et al. 1989). Numerical simulations which directly tackle  the 
decoupling issue in this regime have been very limited so far. Friedli \&
Martinet (1993, and priv. communication) also experienced difficulties in decoupling
pure stellar disks. As mentioned above, the inner stellar bar in this case is
very transient and quickly dissappears, although this may be a result of
an insuffient number of particles. Combes (1994) was unable to obtain decoupling
in the purely stellar {\sl and} gaseous cases. For a mixed system of gas and
stars, the decoupled stage is more prolonged. Hence, numerical simulations
clearly favor  decoupling  in the non-linear stage and explicitly
demonstrate the necessity for  gas to be present, in addition to the
stellar component.          

Numerical simulations also confirm that an increased central-mass concentration
is important for the dynamical separation of the outer and inner parts. This can
be only achieved by moving the gas by means of a large-scale stellar bar. The
gas accumulates inside the ILR in the $x_2$ orbits, modifying the local
potential and ultimately forming a double ILR. The gravity of the gas settling
into these orbits is sufficient to ``drag'' the stars along and twist the NIR
isophotes (e.g., Shaw et al. 1993), but such a configuration still corotates with
the stellar bar. The ability of the gas to settle into the
 $x_2$ orbits depends upon
its sound speed and viscosity. When the gas is too viscous or too hot, it will
avoid the ILRs completely and remain in the $x_1$ orbits aligned with the bar
(Englmaier \& Gerhard 1997; Englmaier \& Shlosman 2000; Patsis \&
Athanassoula 2000). Moderately viscous gas will settle into the innermost $x_2$
orbits, losing  angular momentum due to  viscous torques in addition to
the gravitational torques from the bar. The 
evolution of nuclear regions in disks,
therefore, depends on the (unknown) equation of state of the ISM. 
The inclusion of star
formation in  nuclear rings has already demonstrated how the resulting increase
in viscosity leads to the mass transfer across the ILRs, as evident from
Figure~2. 

Alternatively, Englmaier \& Shlosman (1999, unpublished) have modified the
ZEUS-2D hydrocode (Stone \& Norman 1992) to study the stability and evolution
of nuclear rings discussed in \S 4.1, in the presence of
gravitational effects in the gas. In the first stage, the nuclear rings
started to librate around the potential valley of the stellar bar. However,
almost immediately the rapid growth of self-gravitating modes with $m=2$ and 4
has led to the appearance of a prograde mode (i.e., in the direction of gas
rotation) with pattern speed much larger than $\Omega_{\rm p}$. This has resulted in
an avalanche-type flow towards smaller radii. A model with the central SBH
revealed that inflowing gas feeds the SBH at peak rates, increasing its mass
tenfold. The exact analysis of unstable global modes in these configurations
is under way, in particular their relationship to $I$ and $J$ modes found in
Christodoulou \& Narayan (1992) and Christodoulou (1993), and to stability
of the Maclaurin sequence of self-gravitating rotating fluids (Christodoulou 
et al. 1995). 

A big unknown is the concurrent
star formation, which was neglected in these simulations. Note, however, that
the star formation has low efficiency and can hardly halt this runaway
collapse toward the center. 

\section{Final Remarks}

The central kpc of disk galaxies can host a number of non-axisymmetric
morphological features, from mild perturbations, such as nuclear spirals driven
by large-scale bars, to triaxial bulges, to strong secondary bars, gaseous or
stellar. Their effects on the dynamics of the ISM and its state are complex and
require a long-range effort in order to understand the central activity, which
includes massive star formation and fueling of the central SBHs. New
observations indicate that the state of the cold-phase ISM within the central
kpc may differ from that in the surrounding disk (e.g., H\"uttemeister \&
Aalto, this
volume, p. 617) and numerical simulations reveal little analogy between the gas
dynamics in the central regions and in the outer disks.

New higher-resolution numerical simulations involving much larger amounts of
particles will clarify the issues related to nested bar formation and
self-gravitating gas dynamics in these systems. Coupled with new observational
techniques and instruments at longer wavelengths, they will shed light
on the intricate evolution of central regions within the general context
of cosmological evolution.

\acknowledgements I am grateful to my collaborators Peter Englmaier, Clayton Heller, Johan
Knapen, Seppo Laine, and Reynier Peletier. I thank the organizers of this conference
for financial assistance. This work is 
supported in part by NASA grants NAG 5-10823, HST
GO-08123.01-97A, and WKU-522762-98-6.

\section*{Discussion}

\noindent {\it Beckman:\/} So is the object we have just seen a bar or a ring?

\medskip

\noindent {\it Shlosman:\/} Gaseous bars are basically very elongated rings.
This is the only way to avoid the intersecting orbits which gas can{\sl not}
populate. In the case of intersecting orbits, this would limit the bar lifetime
to a fraction of a rotation period.

\medskip

\noindent {\it Beckman:\/} Your movie showed the inner bar with a
{\sl slower} pattern speed than the main bar, but are the observations
not showing that the nuclear bars have higher pattern speeds?

\medskip

\noindent {\it Shlosman:\/}  Pattern speeds of nuclear bars, either stellar
or gaseous, have not yet been determined. We have presented for the first
time an instability in a gaseous ring which leads to a pattern speed
slower than that of the primary stellar bar.

\medskip

\noindent {\it Elmegreen:\/} Why do the stars that are born in a gaseous
inner ring stay in the ring long after star formation is over and the gas
is dispersed (as we heard from Peter Erwin this morning)? This implies
that gas dynamical processes do not define the equilibrium gaseous ring
position, but only gravity does.

\medskip

\noindent {\it Shlosman:\/} This is hardly surprising. There are a number of
possibilities for removing the gas from the original position of the nuclear
ring. First, the bar instability in a weakly self-gravitating gas shown in
our work removes a sufficient amount of angular momentum and energy (i.e.,
Jacobi energy) from the gas transferring it deeper into the potential well.
If stars coexisted initially with the gas, they would go through the same
process. In other words, stars stay where they form at
 their original (Jacobi) energies. We have checked the dynamical
evolution of such a stellar ring  and found that it thickens
initially due to the differential precession of individual stellar orbits,
and then any dynamical evolution slows down dramatically. So the red nuclear
stellar rings mentioned before (Shlosman 1999) are a normal product
of the cut-off of the gas supply to the ring (e.g., due to bar weakening
or the absence of available gas).

\medskip

\noindent {\it Norman:\/} Does the final state of the inner bar
have the same pattern speed as the outer primary bar?

\medskip

\noindent {\it Shlosman:\/}  The dynamical decoupling is always limited in time
and to particular cases (stellar, gaseous, or mixed decouplings). The 
instability discussed here, leading to the decoupling of a 
low-surface-density  (weakly self-gravitating) gas results in pattern speeds
that are {\sl lower} by a factor of 2--3 than the pattern speed of the
primary stellar bar.


\begin{references}

\reference Athanassoula, E. 1992, \mnras, 259, 345

\reference Binney, J., \& Tremaine, S. 1987, Galactic Dynamics (Princeton: Princeton University
   Press)

 
\reference Buta, R., \& Crocker, D. A. 1993, \aj, 105, 1344  

\reference Christodoulou, D. M. 1993, \apj, 412, 696

\reference Christodoulou, D. M., Kazanas, D., Shlosman, I., \& Tohline, J. E. 1995,
     \apj, 446, 510  

\reference Christodoulou, D. M., \& Narayan, R. 1992, \apj, 388, 451   
 

\reference Combes, F. 1994, { Mass-Transfer Induced Activity in Galaxies,}
     ed. I. Shlosman (Cambridge: Cambridge University Press), p. 170   

\reference Contopoulos, G., \& Papayannopoulos, T. 1980, A\&A, 92, 33

\reference Davies, C. L., \& Hunter J. H., Jr. 1997, \apj, 484, 79 

\reference de Vaucouleurs, G. 1974, in IAU Symp. 58, The Formation and
   Dynamics of Galaxies, ed. J.R. Shakeshaft (Dordrecht: Reidel), p. 335 

\reference de Vaucouleurs, G., de Vaucouleurs, A., Corwin H. G., Jr., Buta, R.,
   J., Paturel, G., \& Fouque, P. 1991, 3rd Reference Catalogue of
   Bright Galaxies (New York: Springer)  

\reference Devereux, N. A., Kenney, J. D. P., \& Young, J. S. 1992, \aj, 103, 784 


\reference Elmegreen, B. G. 1994, \apj, 425, L73 

\reference Elmegreen, B. G., \& Elmegreen, D. M. 1985, \apj, 288, 438  

\reference  Elmegreen, D. M., Chromey, F. R., \& Santos, M. 1998, \aj, 116,
    1221 

\reference Emsellem, E., Greusard, D., Combes, F., Friedli, D., Leon, S.,
    P\'econtal, E., \& Wozniak, H. 2001, A\&A, 368, 52

\reference Englmaier, P., \& Gerhard, O. 1997, \mnras, 287, 57

\reference Englmaier, P., \& Shlosman, I. 2000, \apj, 528, 677


\reference Erwin, P., \& Sparke, L. S. 1999, \apj, 521, L37 

\reference Friedli, D. 1999, ASP Conf. Ser., vol. 187,
{Evolution of Galaxies on Cosmological
     Timescales,} ed. J. E. Beckman \& T. J.~Mahoney, eds. (San Francisco: ASP).
p. 88 

  \reference Friedli, D., \& Martinet, L. 1993, A\&A, 277, 2   

\reference Friedli, D., Wozniak, H., Rieke, M., Martinet, L., \& Bratschi, P.
     1996, A\&AS, 118, 461  


\reference Heller, C. H., \& Shlosman, I. 1994, \apj, 424, 84

\reference Heller, C. H., \& Shlosman, I. 1996, \apj, 471, 143

\reference Heller, C. H.,  Shlosman, I., \& Englmaier, P. 2001, \apj, 553, 661

\reference Ishizuki, S., Kawabe, R., Ishiguro, M., Okumura, S. K., \& Morita,
   K.-I. 1990, Nat, 344, 224 
 
\reference Jogee, S., Kenney, J. D. P., \& Smith, B. J. 1998,
   \apjl, 494, L185

\reference Jungwiert, B., Combes, F., \& Axon, D. J. 1997, A\&AS, 125, 479  

 \reference Knapen, J. H., Beckman, J. E., Heller, C. H., Shlosman, I., \& de Jong,
   R. S. 1995, \apj, 4, 623
 
\reference Knapen, J. H., Shlosman, I., \& Peletier, R. F. 2000, \apj, 529, 93
                                                                            

\reference Kormendy, J. 1982, \apj, 257, 75

\reference Kotilainen, J. K., Reunanen, J., Laine, S., \& Ryder, S. D. 2000, A\&A,
   353, 834 

\reference Laine, S., Knapen, J. H., Perez-Ramirez, D., Doyon, R., \& Nadeau, D.
   1998, \mnras, 302, 33L


\reference Laine, S., Shlosman, I., Knapen, J. H., \& Peletier, R. F. 2001, \apj,   in press (astro-ph/0108029)

\reference Maiolino, R., Alonso-Herrero, A., Anders, S., Quillen, A., Rieke,
   M. J., Rieke, G. H., \& Tacconi-Garman, L. E. 2000, \apj, 531, 219

\reference  Martini, P., \& Pogge, R. W. 1999, \aj, 118, 2646
 
\reference Mulchaey, J. S., \& Regan, M. W. 1997, \apjl, 482, L135  

\reference Patsis, P. A., \& Athanassoula, E. 2000, \mnras, 358, 45    

\reference Pfenniger, D., \& Norman, C. A. 1990, \apj, 363, 391 

\reference Prendergast, K. H. 1962, Distribution and Motion of ISM in Galaxies,
   ed. L. Woltjer (New York: Benjamin), p. 217

\reference Regan, M. W., \& Mulchaey, J. S. 1999, \aj, 117, 2676  

\reference Sandage, A., \& Brucato, R. 1979, \aj, 84, 472

\reference Sellwood, J. A., \& Merritt, D. 1994, \apj, 425, 530   

\reference Sellwood, J. A., \& Moore, E. M. 1999, \apj, 510, 125     

\reference Shaw, M. A., Axon, D. J., Probst, R., \& Gatley, I. 1995, \mnras,
     274, 369

\reference Shaw, M. A., Combes, F., Axon, D. J., \& Wright, G. S. 1993,
     A\&A, 273, 31   
  
\reference Shlosman, I. 1999, ASP Conf. Ser., vol. 187, Evolution of Galaxies on Cosmological
     Timescales, ed. J. E. Beckman \& T. J.~Mahoney,  (San Francisco: ASP),
p.  100    

\reference Shlosman, I., Frank, J., \& Begelman, M. C. 1989, Nat, 338, 45 

\reference Shlosman, I., \& Heller, C. H. 2001, \apj, submitted
 
\reference Stone, J. M., \& Norman, M. L. 1992, \apjs, 80, 753  

\end{references}
\end{document}